# Security theory for data flow and access control:
## From partial orders to lattices and back, a half-century trip

Luigi Logrippo[1,2][0000-0001-8804-0450]

[1] Université du Québec en Outaouais, Département d'informatique et ingénierie
[2] University of Ottawa, School of Information Technology and Engineering
luigi@uqo.ca

**Abstract.** The multi-level Bell-La Padula model for secure data access and data flow control, formulated in the 1970's, was based on the theory of partial orders. Since then, another model, based on lattice theory, has prevailed. We present reasons why the partial order model is more appropriate. We also show, by example, how non-lattice data flow networks can be easily implemented by using Attribute-based access control (ABAC).

## 1   Introduction

This is a *position paper* in the framework of the theory of data security in *networks* (also called *data flows*). It presents discussion and background complementing some previous papers, and is not self-contained. In data flow security theory, one considers communicating *entities* representing users, subjects, data objects such as databases, files, etc. The security problem is to ensure that data originating from certain entities can end up in specific other entities only, by direct or indirect data transfer. Since data carry information, constraining the flow of data implies constraining the flow of information.

Access control rules among entities define *Channels*, through which data can be directly transmitted. But, through chains of channels, data can also be transmitted indirectly to other entities, including entities that may not have direct access to them. This possibility is represented by the transitive, reflexive closure of the Channel relation, the *CanFlow* relation. The CanFlow relation is a preorder, which can contain equivalence classes of entities that can get the same data, directly or indirectly. By taking such equivalence classes as units, the result is a partial order of equivalence classes [Fraïssé 1986]. According to the position of their equivalence classes in the partial order, entities can be assigned *security labels* that can be used by access control rules to define channels and data flows. Partial orders and labels define *Multi-level systems (MLS),* which can control both direct and indirect data access.

This theory, developed starting in the 1970's for operating system and database security, has evolved slowly, but is still relevant today for data security and privacy in the Cloud, the IoT, and future-generation networks.

## 2      Historical background and discussion

The first substantial research reports on the theory of data flow control for security were the well-known MITRE Corporation technical reports of David E. Bell and Len J. La Padula, produced over the years 1972-76. Reference [Bell, La Padula 1976] (BLP henceforth) has an Appendix that presents a formal model, based on a partial order relation among labels. The report does not mention lattices and the example it provides (its Fig. A1) shows a partial order of labels that is clearly not a lattice. The BLP model has been criticized as being too rigid, however all access control models define partial orders of labels, just as the BLP model does [Logrippo 2021]. It is unfortunate that most examples relating to it show only totally ordered sets of labels. In fact, our Example of Fig. 1 is consistent with the BLP model, and the split entities that we show, which belong to two data flows but keep them separate, correspond to BLP's *trusted entities* (data flows crossing such entities have been called *intransitive*).

In the same years, [Denning 1976] proposed that the partial order among security labels must be a *lattice*, which means that any two labels must have unique joins and meets. This is a strong requirement. A companion paper [Denning, Denning, Graham 1976] starts with the assertion *"it has been shown that lattice-structured policies[1] have properties which lead to simple and efficient enforcement mechanisms"*. It cites as evidence several papers, including some that mention only notions of order, not lattices. It shows that non-lattice structured *policies* can be transformed into lattices *"while preserving the validity of all flows"*, by a method involving the definition of new labels, not assigned to entities, which we will call *void labels*. The main example (their Fig. 1c) shows a partial order of labels, called *initial policy*, which can be implemented by using the method exemplified in Section 3 below. However the authors are not satisfied with this and go on to show that by adding four void labels with channels leading to and from them, this partial order can be transformed into a lattice. But, ***are the unique joins and meets, or the added void labels and channels, useful for data security?*** Consider the following:

1. In many organizations, the void labels might contradict explicit security policies. For example, they may include a label for entities that have access to all data, as well as a label for entities that have access to none [Denning, Denning, Graham 1976, Sandhu 1993]. Also, for any two labels, there must be a label for entities that can receive data from both, in possible contradiction of conflict-of-interest policies.
2. As recognized in [Denning, Denning, Graham 1976], changes (or reconfigurations) in the network structure may be needed frequently: consider VANET networks, where hundreds of entities can be involved, with many reconfigurations per second. The unnecessary algorithm for recovering the lattice structure of the set of labels will have to be executed at each reconfiguration, with consequent addition of new void labels.

---

[1] The *policies* in their theory are called *networks* or *data flows* in ours. We use the term *policy* in the broader sense common in security. Policies determine data flows.

3. Since a subset of a lattice is not necessarily a lattice, the fact that a lattice data flow policy is implemented in an organization does not necessarily imply that it is implemented in all parts of the organization.
4. Since the union of two lattices is not a lattice, when two organizations merge, it will be necessary to add void labels and new channels in order to obtain a lattice of labels. Even if the data flow in each organization is secure, the data flow in their union would not be secure until these labels are added.

Almost all subsequent research on data flow control for security, including textbooks and recent papers, has been influenced by the 'lattice' model, occasionally softened in terms of 'semi-lattices'. We find many explicit or implicit statements that a lattice structure of labels is necessary for secure data flow control, without giving reasons. BLP and related MLS were considered to be lattice models, and this was confirmed by D.E. Bell [Bell 1990], apparently without considering that this resulted in a restriction of the generality of the models. Several authors have noted that RBAC can define data flows that form partial orders but not lattices, hence the view that RBAC is more powerful than BLP or MLS. With the partial order original model for BLP, they have the same power [Logrippo 2025].

[Sandhu 1993] elaborates on the use and properties of the lattice model. It claims that the BLP model is lattice-based. It does express doubts about the lattice model, however. Sandhu writes that *"although this article focuses on policies that satisfy Denning's axioms, there are legitimate information flow policies that do not satisfy these axioms"*. Also he shows an example illustrating that *"in some situations it might be more appropriate to use partially ordered labels than to strive for a complete lattice"*.

Indeed, in support of these doubts, by showing that any network has an implicit partial order [Logrippo 2021], it can be proved that:

A. No void labels need to exist since labels can be assigned using any network's partial order structure: simply, each equivalence class of entities gets its unique label by a bottom-up construction, following the CanFlow relation that determines label inclusion (Section 3 below). Thus, any network of communicating entities defines a partially ordered data flow of labels and entities, from equivalence classes that cannot receive from any others (thus having minimal labels and maximum integrity) to equivalence that cannot send to any others (thus having maximal labels and maximum secrecy) [Logrippo 2021, Logrippo 2024].
B. Changes or reconfigurations in networks lead from partial orders to partial orders, and no void labels need to be added. If the partial order that a network defines does not conform to the security policy that the network is supposed to implement, the network's partial order can be reconfigured to conform [Logrippo 2021].
C. Subsets of partial orders are partial orders. Any network and all its parts implement a data flow security policy by which the data flow is determined by label inclusion.

   D.   Unions of partial orders are partial orders, so, in the case of union of organizations, no void labels have to be added in order to achieve data security.

Point C can be explained in terms of the policy of an organization where entities have labels, stating what data they can receive, and there is a channel (or data flow) from entity *x* to entity *y* iff the label of *x* is included in the label of *y*. The decision function of all access control and data flow control models can be defined in terms of this policy (see Section 3). The differences among models lie in the formalisms used to express the policies and the reconfigurations, or the administration.

## 3     Example with labels as attributes in ABAC

In a company, we have the following entities:
1) Four points of sales: S1, S2, S3, S4.
2) Three data processing centers: P1,P2,P3.
3) Two centers for data analysis, A1, A2.
4) A center to produce orders, O.

The security policies require the existence of two separate data flows, one for sales data, and the other for statistical data. For the first data flow, the (informally specified) rules are:
1) Sales data coming from S1 go to P1
2) Sales data coming from S2 go to P2
3) Sales data coming from S3 and S4 go to P3
4) P1 and P2 can share data freely
5) Data from P3 goes to P2
6) Data from P2 goes to A1 and A2
7) A1 and A2 cannot get the results of each other's processing.

In this data flow, note the equivalence class {P1,P2}(rule 4), which constitutes a single element in the partial order. Although P1 and P2 have different channels, what matters from the security point of view is that the same data can flow to both of them, directly or indirectly. Note also the 'conflicts' among entities that cannot exchange data, neither directly, nor indirectly. Except for 7), these are implicit, assuming a 'closure' policy stating that the data flows that are not explicitly permitted are forbidden.

In the second data flow, we assume the existence of a trusted program that compiles statistics from the data of A1, including stocking levels and excluding identification data such as names of sales points. The results are given to a separate part of A1, called A1S, which sends them to a separate part of A2, called A2S, and to O. So this is a separate data flow, that starts in A1S, involving different data. Examples involving such policies, including data separation, have been reported in the literature, see for example [Myers and Liskov 1997 and 2000]. We give this data flow only one rule:
   1') Data in A1S go to A2S and O.

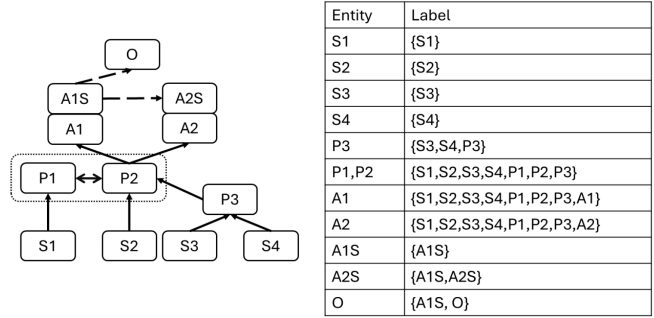

**Figure 1a)** Two data flows with **1b)** their set of labels

These two policies are shown, in the form of a channel configuration, in the graph of Fig. 1a). The first policy is given with continuous arrows, the second with dashed arrows. The arrows represent the channels, and the paths in the graph with the same types of arrows give the CanFlow relation. That is, data from S1 cannot flow to O, it is a different type of data that can flow from A1S to O. This represents the fact that the second flow has been cleaned of data that we don't want to reach O. Neither partial order constitutes a lattice and making them into lattices would require the addition of extraneous entities and channels, e.g. adding a join for A1 and A2 implies adding an entity that can know the data of both entities, in violation of the company's policy. However a set of labels can be calculated, as shown in Fig. 1b). As defined in [Logrippo 2021], each entity's label is taken to be the set of the names of the entities it dominates in the partial order. In other words, the label of each entity denotes the data that can reach the entity, their *provenance*, see Fig. 1b). Entities in the same equivalence class have the same label. We have a Multi-level system (MLS) as usually described in the data security literature.

The implementation can be done in several ways. In [Stambouli, Logrippo 2024] we have shown how labels can be used to direct data flows in Software Defined Networks (SDN). The data flows can also be implemented by encryption, to block all forbidden data transfers. We show here how the labels can become attributes for an Attribute-based access control (ABAC) [Hu et al. 2015] implementation.

The ABAC system to implement the specified flow can be constructed as follows:
1. The labels are used as attributes of the entities. The label table is stored in the Policy Information Point.
2. The policy in the Policy Administration Point, to be evaluated by the Policy Decision Point, is the following: *a request for an operation that moves data from an entity* x *to an entity* y *is granted iff the label of* x *is included in the label of* y.

Whether an operation 'moves data' is a matter of practical considerations to be considered by security specialists. There are channels wherever such operations are allowed.

Note for example that, according to this policy, the data of A1 cannot flow to A2 or S1, directly or indirectly. The data of S1 cannot flow to A1S or O (they need to be compiled into statistics first). Note also that this policy actually implements the transitive closure of the channel configuration of Fig. 1. For example, according to it, there is a channel from S1 to A1, although this is not mentioned in the initial requirements, nor shown in the figure. From the point of view of data security, these additional channels do not make a difference, since A1 can get the data of S1 indirectly through P1 and P2. We take a pessimistic view, by which any allowed data transfer can occur, and this is consistent with the literature, starting with the historical papers mentioned above.

Since this method can be used to implement any data flow that can be represented as an order relation such as the one of Fig. 1a), it should be clear that a partial order structure is sufficient to define and implement any data security network, whether it is lattice-formed or not.

If the network is reconfigured, the labels have to be recomputed, but this is an efficient process [Stambouli, Logrippo 2019]. In realistic systems, the labels can be very long, but this can be addressed by giving names to sets of labels, in the same way that a single name, such as "Classified" can be used to identify large numbers of documents and users.

## 4      Conclusions

With respect to partial order data flow theory, lattice data flow theory has limitations that are not compensated for by advantages, apart from specific applications where the existence of unique joins and meets may be necessary. Going back to the original justification for the lattice model, we have shown above that partial order-structured policies, in general, can be implemented by simple and efficient access control mechanisms. We must go back to the partial order model proposed fifty years ago by Bell and La Padula, and explore the possibilities of partial order theory in its full generality and many specializations, in the recent application areas of data security and privacy protection in the IoT, the Cloud, mobile networks, etc.

**Acknowledgment.** This research has been funded in part by the Natural Sciences and Engineering Research Council of Canada under their Discovery Grant program.